\RequirePackage{lineno}
\documentclass[aps,prl,twocolumn,showpacs,superscriptaddress,groupedaddress]{revtex4}  
\usepackage{graphicx}  
\usepackage{dcolumn}   
\usepackage{bm}        
\usepackage{amssymb}   
\usepackage{lineno}
\usepackage{url} 
\usepackage{xcolor}

\begin{document}

\title{Hard exclusive pion electroproduction at backward angles with CLAS}

\newcommand*{\ANL}{Argonne National Laboratory, Argonne, Illinois 60439, USA}
\newcommand*{\ANLindex}{1}
\affiliation{\ANL}
\newcommand*{\CSUDH}{California State University, Dominguez Hills, Carson, CA 90747, USA}
\newcommand*{\CSUDHindex}{2}
\affiliation{\CSUDH}
\newcommand*{\CMU}{Carnegie Mellon University, Pittsburgh, Pennsylvania 15213, USA}
\newcommand*{\CMUindex}{3}
\affiliation{\CMU}
\newcommand*{\CUA}{Catholic University of America, Washington, D.C. 20064, USA}
\newcommand*{\CUAindex}{4}
\affiliation{\CUA}
\newcommand*{\SACLAY}{IRFU, CEA, Universit\'e Paris-Saclay, F-91191 Gif-sur-Yvette, France}
\newcommand*{\SACLAYindex}{5}
\affiliation{\SACLAY}
\newcommand*{\CNU}{Christopher Newport University, Newport News, Virginia 23606, USA}
\newcommand*{\CNUindex}{6}
\affiliation{\CNU}
\newcommand*{\UCONN}{University of Connecticut, Storrs, Connecticut 06269, USA}
\newcommand*{\UCONNindex}{7}
\affiliation{\UCONN}
\newcommand*{\FU}{Fairfield University, Fairfield CT 06824, USA}
\newcommand*{\FUindex}{8}
\affiliation{\FU}
\newcommand*{\FERRARAU}{Universita' di Ferrara , 44121 Ferrara, Italy}
\newcommand*{\FERRARAUindex}{9}
\affiliation{\FERRARAU}
\newcommand*{\FIU}{Florida International University, Miami, Florida 33199, USA}
\newcommand*{\FIUindex}{10}
\affiliation{\FIU}
\newcommand*{\FSU}{Florida State University, Tallahassee, Florida 32306, USA}
\newcommand*{\FSUindex}{11}
\affiliation{\FSU}
\newcommand*{\Genova}{Universit$\grave{a}$ di Genova, 16146 Genova, Italy}
\newcommand*{\Genovaindex}{12}
\affiliation{\Genova}
\newcommand*{\GWUI}{The George Washington University, Washington, DC 20052, USA}
\newcommand*{\GWUIindex}{13}
\affiliation{\GWUI}
\newcommand*{\ISU}{Idaho State University, Pocatello, Idaho 83209, USA}
\newcommand*{\ISUindex}{14}
\affiliation{\ISU}
\newcommand*{\INFNFE}{INFN, Sezione di Ferrara, 44100 Ferrara, Italy}
\newcommand*{\INFNFEindex}{15}
\affiliation{\INFNFE}
\newcommand*{\INFNFR}{INFN, Laboratori Nazionali di Frascati, 00044 Frascati, Italy}
\newcommand*{\INFNFRindex}{16}
\affiliation{\INFNFR}
\newcommand*{\INFNGE}{INFN, Sezione di Genova, 16146 Genova, Italy}
\newcommand*{\INFNGEindex}{17}
\affiliation{\INFNGE}
\newcommand*{\INFNRO}{INFN, Sezione di Roma Tor Vergata, 00133 Rome, Italy}
\newcommand*{\INFNROindex}{18}
\affiliation{\INFNRO}
\newcommand*{\INFNTUR}{INFN, Sezione di Torino, 10125 Torino, Italy}
\newcommand*{\INFNTURindex}{19}
\affiliation{\INFNTUR}
\newcommand*{\ORSAY}{Institut de Physique Nucl\'eaire, CNRS/IN2P3 and Universit\'e Paris Sud, Orsay, France}
\newcommand*{\ORSAYindex}{20}
\affiliation{\ORSAY}
\newcommand*{\ITEP}{Institute of Theoretical and Experimental Physics, Moscow, 117259, Russia}
\newcommand*{\ITEPindex}{21}
\affiliation{\ITEP}
\newcommand*{\JMU}{James Madison University, Harrisonburg, Virginia 22807, USA}
\newcommand*{\JMUindex}{22}
\affiliation{\JMU}
\newcommand*{\KNU}{Kyungpook National University, Daegu 41566, Republic of Korea}
\newcommand*{\KNUindex}{23}
\affiliation{\KNU}
\newcommand*{\MISS}{Mississippi State University, Mississippi State, MS 39762-5167, USA}
\newcommand*{\MISSindex}{24}
\affiliation{\MISS}
\newcommand*{\UNH}{University of New Hampshire, Durham, New Hampshire 03824-3568, USA}
\newcommand*{\UNHindex}{25}
\affiliation{\UNH}
\newcommand*{\NSU}{Norfolk State University, Norfolk, Virginia 23504, USA}
\newcommand*{\NSUindex}{26}
\affiliation{\NSU}
\newcommand*{\OHIOU}{Ohio University, Athens, Ohio  45701, USA}
\newcommand*{\OHIOUindex}{27}
\affiliation{\OHIOU}
\newcommand*{\ODU}{Old Dominion University, Norfolk, Virginia 23529, USA}
\newcommand*{\ODUindex}{28}
\affiliation{\ODU}
\newcommand*{\RPI}{Rensselaer Polytechnic Institute, Troy, New York 12180-3590, USA}
\newcommand*{\RPIindex}{29}
\affiliation{\RPI}
\newcommand*{\URICH}{University of Richmond, Richmond, Virginia 23173, USA}
\newcommand*{\URICHindex}{30}
\affiliation{\URICH}
\newcommand*{\ROMAII}{Universita' di Roma Tor Vergata, 00133 Rome Italy}
\newcommand*{\ROMAIIindex}{31}
\affiliation{\ROMAII}
\newcommand*{\MSU}{Skobeltsyn Institute of Nuclear Physics, Lomonosov Moscow State University, 119234 Moscow, Russia}
\newcommand*{\MSUindex}{32}
\affiliation{\MSU}
\newcommand*{\SCAROLINA}{University of South Carolina, Columbia, South Carolina 29208, USA}
\newcommand*{\SCAROLINAindex}{33}
\affiliation{\SCAROLINA}
\newcommand*{\TEMPLE}{Temple University,  Philadelphia, PA 19122, USA}
\newcommand*{\TEMPLEindex}{34}
\affiliation{\TEMPLE}
\newcommand*{\JLAB}{Thomas Jefferson National Accelerator Facility, Newport News, Virginia 23606, USA}
\newcommand*{\JLABindex}{35}
\affiliation{\JLAB}
\newcommand*{\UTFSM}{Universidad T\'{e}cnica Federico Santa Mar\'{i}a, Casilla 110-V Valpara\'{i}so, Chile}
\newcommand*{\UTFSMindex}{36}
\affiliation{\UTFSM}
\newcommand*{\EDINBURGH}{Edinburgh University, Edinburgh EH9 3JZ, United Kingdom}
\newcommand*{\EDINBURGHindex}{37}
\affiliation{\EDINBURGH}
\newcommand*{\GLASGOW}{University of Glasgow, Glasgow G12 8QQ, United Kingdom}
\newcommand*{\GLASGOWindex}{38}
\affiliation{\GLASGOW}
\newcommand*{\VT}{Virginia Tech, Blacksburg, Virginia   24061-0435, USA}
\newcommand*{\VTindex}{39}
\affiliation{\VT}
\newcommand*{\VIRGINIA}{University of Virginia, Charlottesville, Virginia 22901, USA}
\newcommand*{\VIRGINIAindex}{40}
\affiliation{\VIRGINIA}
\newcommand*{\WM}{College of William and Mary, Williamsburg, Virginia 23187-8795, USA}
\newcommand*{\WMindex}{41}
\affiliation{\WM}
\newcommand*{\YEREVAN}{Yerevan Physics Institute, 375036 Yerevan, Armenia}
\newcommand*{\YEREVANindex}{42}
\affiliation{\YEREVAN}
 
\newcommand*{\Ecole}{Centre de physique th\'eorique, {\'E}cole Polytechnique, CNRS, F-91128 Palaiseau, France}
\newcommand*{\Ecoleindex}{43}
\affiliation{\Ecole}
\newcommand*{\PNPI}{National Research Centre Kurchatov Institute, Petersburg Nuclear Physics Institute, RU-188300 Gatchina, Russia}
\newcommand*{\PNPIindex}{44}
\affiliation{\PNPI}

\newcommand*{\NOWISU}{Idaho State University, Pocatello, Idaho 83209, USA}
\newcommand*{\NOWJLAB}{Thomas Jefferson National Accelerator Facility, Newport News, Virginia 23606, USA}
\newcommand*{\NOWGLASGOW}{University of Glasgow, Glasgow G12 8QQ, United Kingdom}
\newcommand*{\NOWINFNGE}{INFN, Sezione di Genova, 16146 Genova, Italy}
\author {K.~Park} 
\affiliation{\JLAB}
\author {M.~Guidal} 
\affiliation{\ORSAY}
\author {R.W.~Gothe} 
\affiliation{\SCAROLINA}

\author {B.~Pire} 
\affiliation{\Ecole}
\author {K.~Semenov-Tian-Shansky} 
\affiliation{\PNPI}
\author {J.-M.~Laget} 
\affiliation{\JLAB}

\author {K.P. ~Adhikari} 
\affiliation{\MISS}
\author {S. Adhikari} 
\affiliation{\FIU}
\author {Z.~Akbar} 
\affiliation{\FSU}
\author {H.~Avakian} 
\affiliation{\JLAB}
% ...

\collaboration{The CLAS Collaboration}
\noaffiliation


\date{\today}

\begin{abstract}
We report on the first measurement of cross sections for exclusive
deeply virtual pion electroproduction off the proton, $e p \to e^\prime n \pi^+$, above the resonance region
at backward pion center-of-mass angles.
The $\varphi^*_{\pi}$-dependent cross sections
were measured, from which we extracted three
combinations of structure functions of the proton.
Our results are compatible with calculations based on nucleon-to-pion transition distribution amplitudes (TDAs) and shed new light on nucleon structure.
\end{abstract}

\modulolinenumbers[1]
\maketitle


During the past two decades the study of hard exclusive processes
has significantly increased the understanding of hadron structure
in terms of the fundamental degrees of freedom of
Quantum Chromo-Dynamics (QCD), the quarks and gluons.
The QCD collinear factorization theorems state that for
special kinematic conditions a broad class of hard exclusive reactions
can be described in terms of universal nucleon structure functions that
 depend on variables
such as the parton longitudinal momentum fractions and impact parameter, which encode the
complex quark and gluon structure of hadrons.
Deeply Virtual Compton Scattering (DVCS) off nucleons
($e N \to e^\prime N^\prime \gamma$)
and hard exclusive electroproduction of mesons off nucleons
($e N \to e^\prime N^\prime M$) in the generalized Bjorken limit
(sufficiently large lepton momentum
transfer squared $Q^2$
and center-of-mass energy squared
$W^2=m_p^2 + 2m_p\nu - Q^2$ for fixed Bjorken
$x_{BJ}=Q^2/(W^2+Q^2-m_p^2)$ and small nucleon momentum transfer
$|t|$) probe the quark and gluon Generalized Parton Distributions (GPDs) of the nucleon.
Here $N$, $N^\prime$, $e$ and $e^\prime$ denote the initial and final nucleon and the initial and final electron, $\nu$ is the electron energy transfer and $m_p$ is the proton mass.

\begin{figure}[!htb]
   \begin{center}
	\includegraphics[angle=0,width=0.24\textwidth]{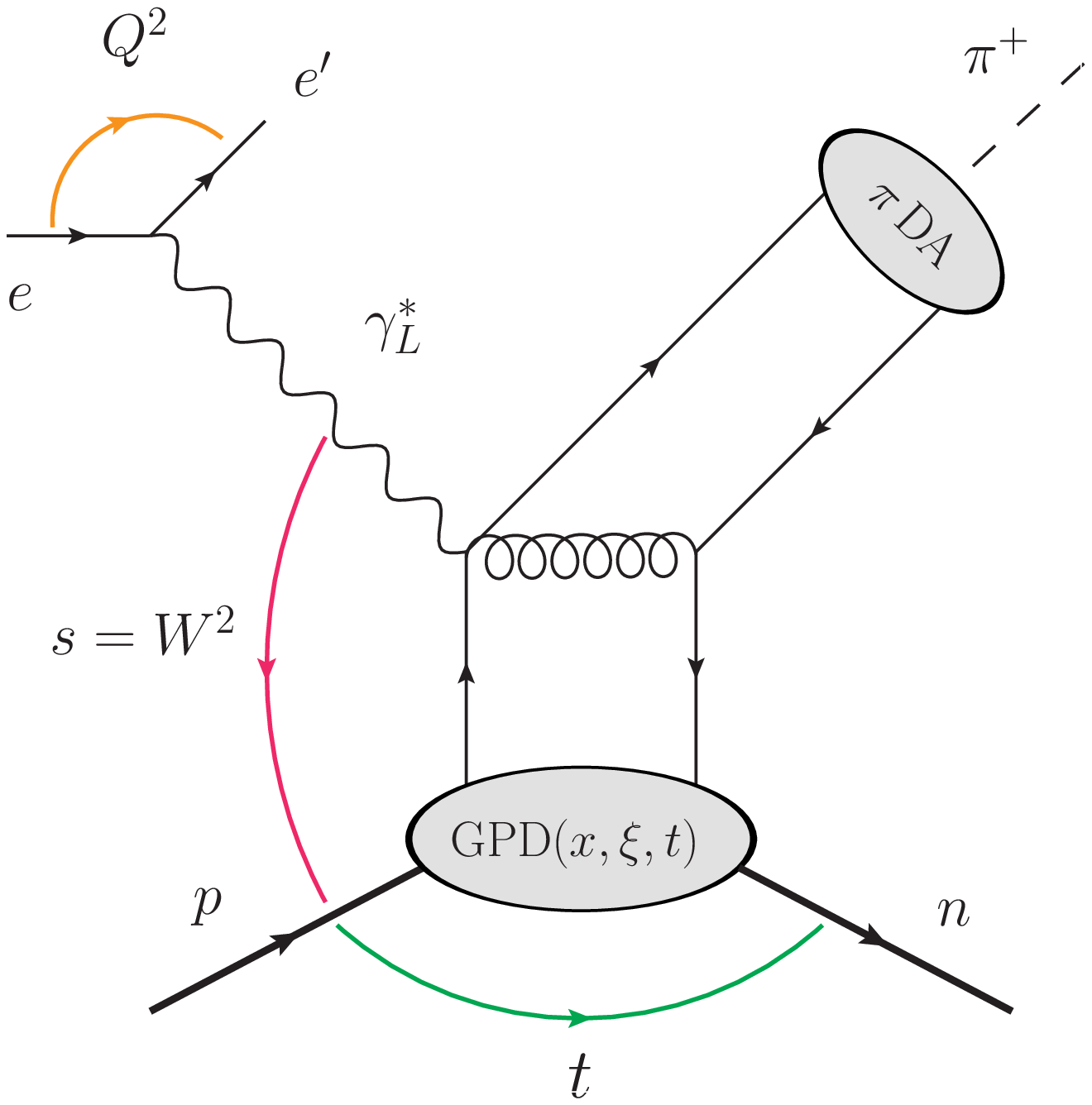}~~~
	\includegraphics[angle=0,width=0.24\textwidth]{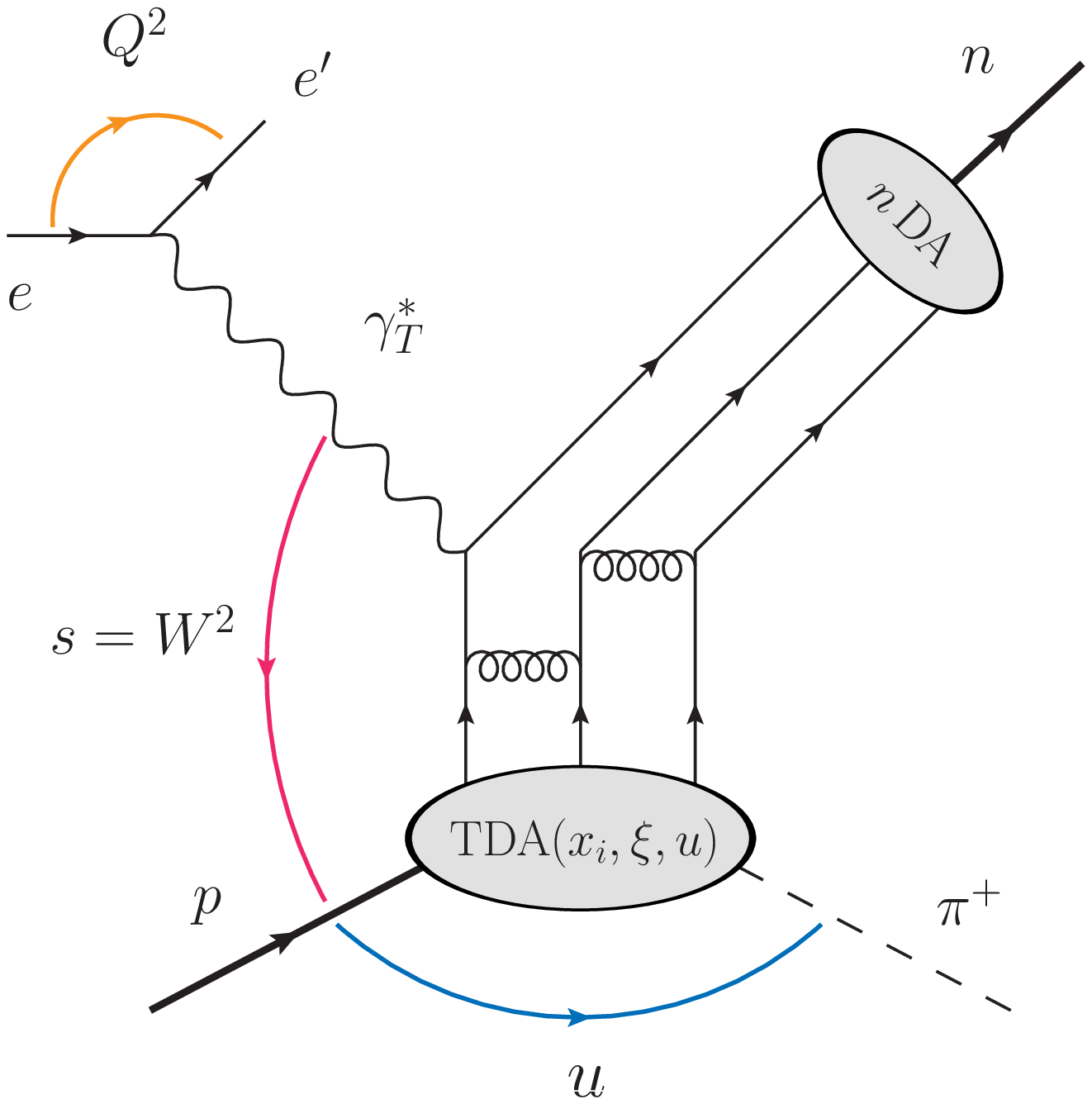}

         \caption[Diagrams]{
        {\bf Left:}
         QCD factorization mechanism for the exclusive electroproduction of a meson ($\pi^+$) on the nucleon (proton) in the ``nearly-forward'' kinematical regime.  At large $Q^2$ and small $|t|$, the amplitude of the process can be presented as a convolution of a hard part  calculable in perturbative QCD and two general structure functions parametrizing the complex non-perturbative structure of the nucleon (the GPDs; bottom blob of the diagram) and of the meson (the pion DA upper blob of the diagram).
{\bf Right:} 
factorization mechanism for the same reaction in the complementary
``nearly-backward'' kinematical regime, where  $Q^2$ and $W^2$ are large, $x_{BJ}$ is fixed
and $|u|$ is kept small.
The amplitude of the process is written as the convolution of the hard interaction amplitude (calculable in perturbative QCD) involving the virtual photon, the three quarks of the out-going nucleon and two gluons, with two structure functions
parametrizing the non-perturbative nucleon-to-pion transitions (TDAs) (bottom blob of the diagram) and the nucleon DA (upper blob of the diagram).
          \label{fig:fg1}
         }
   \end{center}
 \end{figure}

The left panel of
Fig.~\ref{fig:fg1}
illustrates the reaction mechanism involving GPDs
for the
$e p \to e^\prime n \pi^+$
process, which provides information on the correlations
between the longitudinal momentum and transverse spatial distributions of quarks
in the nucleon.
GPDs were also found to be a useful probe of parton
orbital momentum, which contributes to the nucleon spin.
We refer the reader to
Refs.~\cite{Mueller:1998fv,Rady96a,Ji97a,Ji97b}
for the pioneering papers on GPDs and to
Refs.~\cite{Goeke:2001tz,Diehl:2003ny,
Belitsky:2005qn,Boffi:2007yc,Guidal:2013rya,Kumericki:2016ehc}
for reviews of the most important results in the field.
Refs.~\cite{Frankfurt:1999fp,Pire:2005ax,JLansberg} made the case that a collinear factorized description
may be applied to exclusive hard electroproduction of mesons for
the kinematic regime opposite to that 
of GPDs, i.e. the generalized Bjorken limit in which Mandelstam
$|u|$ rather than $|t|$ is small.
In the center-of-mass frame, with the positive
direction chosen along the incoming virtual photon, the small~$|t|$-regime
corresponds to mesons produced in the nearly-forward direction, while
in the small~$|u|$-regime the mesons are produced
in the nearly-backward direction. We will refer to these two distinct
regimes as ``nearly-forward'' and ``nearly-backward'' kinematics.
The universal structure functions accessible in
``nearly-backward'' kinematics are nucleon-to-meson Transition Distribution Amplitudes (TDAs).
On the right panel of~Fig.~\ref{fig:fg1} we illustrate the corresponding factorization
mechanism involving TDAs for $e p \to e^\prime n \pi^+$. In this case, the
non-perturbative part describes a nucleon-meson rather than a nucleon-nucleon transition.
At a fixed QCD factorization scale, 
the nucleon-to-meson TDAs are functions of $x_1$, $x_2$ and $x_3$, the three longitudinal momentum fractions of the quarks
involved in the process, the skewness variable $\xi$ and $u$.
Since momentum conservation imposes the constraint
$\sum_{i} x_i$=$2\xi$, TDAs depend effectively on only $4$ variables.

The information encoded in baryon-to-meson TDAs shares common features with
the nucleon distribution amplitudes (DAs) and the GPDs.
Nucleon-to-meson TDAs characterize partonic correlations inside a nucleon and
provide a tool to study the momentum distribution
of the nucleon's baryon density.
The nucleon-to-meson TDAs involves the same three-quark light-cone operator as the nucleon DA.
However, the TDA is not restricted to the lowest three-quark
Fock state of the nucleon, but is sensitive to $q\bar{q}$-pairs in both the nucleon and meson.
Similar to the GPDs
(see {\it e.g.} Ref.\cite{Dupre:2016mai}),
a Fourier transformed TDA
($\Delta_T \to {\bf b}$)
allows an impact-parameter interpretation
for TDAs in the transverse plane. Depending on the range of $x_i$,
TDAs either describe the process of kicking out a three-quark cluster
from the nucleon at some transverse position
${\bf b}$
or the process of emission of a quark (a pair of quarks) with subsequent reabsorption of
a pair of quarks (a quark) by the final-state meson.
This yields additional information on nucleon structure in the transverse plane and allows femto-photography of hadrons from a new perspective.
We refer the reader to Refs.~\cite{Pire:2011xv,Lansberg:2011aa,Pire:2016aqa}.

In this letter, we present the first experimental results that test the nucleon-to-pion TDA formulation.
We have analyzed for the first time the $e p \to e^\prime n \pi^+$ reaction at relatively large $Q^2$ ($>1.7$ GeV$^2$) and
small $\langle |u| \rangle$ ($=0.5$ GeV$^2$) above the resonance region ($W^2>4$ GeV$^2$), in nearly backward kinematics where the TDA formalism is potentially applicable.
In the one-photon-exchange approximation, the unpolarized exclusive cross section can be factorized as $\sigma(e p \to e^\prime n \pi^+)=\Gamma_v \times \sigma(\gamma^* p \to n \pi^+)$.
The virtual photon flux factor $\Gamma_v$ is given by:
\begin{eqnarray}\label{eq:reaction01}
\Gamma_v=\frac{\alpha_{em}}{2\pi^2}\frac{e^\prime}{e}\frac{W^2-m^2_p}{2m_pQ^2}\frac{1}{1-\epsilon}~,
\end{eqnarray}
where $\alpha_{em}$ is the electromagnetic coupling constant, $\epsilon$ is the virtual photon linear polarization parameter $\epsilon={\left(1+2(\nu^2/Q^2)\tan^2(\theta_e/2)\right)^{-1}}$ and $\theta_e$ is the scattered electron polar angle.
The reduced cross section can then be decomposed as
\begin{eqnarray}\label{eq:reaction02}
\sigma = \sigma_T + \epsilon \sigma_L + \sqrt{2\epsilon(1+\epsilon)}\sigma_{LT} \cos\varphi_\pi^* + \epsilon \sigma_{TT} \cos 2\varphi_\pi^* ~,
\end{eqnarray}
where $\varphi_{\pi}^*$ is the azimuthal angle between the electron scattering plane and the hadronic reaction plane
(the starred variables are understood to be in the virtual photon-proton center-of-mass frame).
The separated cross sections $\sigma_T$, $\sigma_L$,
$\sigma_{LT}$ and $\sigma_{TT}$ depend on $W$, $Q^2$ and $\theta_{\pi}^*$, the polar angle of the $\pi^+$.
The variable $\xi$, on which the TDAs depend, can be approximated as $\xi \sim  {Q^2}/\left({Q^2+2(W^2 + \Delta_T^2 -m_p^2)}\right)$, where $\Delta_T$ is the transverse component of the nucleon-to-pion momentum transfer. The variable $\Delta_T$ can be approximated
by $|p_{\pi}^*| \sin\theta_{\pi}^*$, in which $|p_{\pi}^*|$ is the momentum of the $\pi^+$.
If $Q^2\gg m^2_p$ and $Q^2\gg \Delta_T^2$, then $\xi\approx {x_{BJ}}/{(2-x_{BJ})}$, as in DVCS.
In the calculation of cross sections via the diagram of Fig.~\ref{fig:fg1}-right,
the $x_i$ variables on which the TDAs depend are integrated over and are therefore not directly accessible experimentally.
This is just as the calculation of the cross section of the diagram of Fig.~\ref{fig:fg1}-left
involves an integration over $x$ of the GPDs.


The measurement was carried out with a $5.754$ GeV electron beam energy at Jefferson Lab using the
CEBAF Large Acceptance Spectrometer (CLAS)~{\cite{CLAS}}.
The experimental data were collected with CLAS during the e1-6 run period from October~2001 through January~2002.
CLAS was built around six super-conducting coils arranged
symmetrically in azimuth, generating a toroidal magnetic field around the beam axis.
The six identical sectors of the magnet were independently instrumented with 34 layers of drift chambers (DCs) for charged particle tracking,
plastic scintillation counters for time-of-flight (TOF) measurements, gas threshold Cherenkov counters (CCs) for electron
and pion separation and triggering purposes, and electromagnetic calorimeters (ECs) for photon and neutron detection and electron
triggering. To aid in electron/pion separation, the EC was segmented into an inner part facing the target and an outer part
away from the target. CLAS covered nearly the full $4\pi$ solid angle for the detection of charged particles.
The azimuthal acceptance was maximum at large polar angles and decreased at forward angles.
The e1-6 run had the maximal electron beam energy for the JLab accelerator, which allowed us to
 reach the largest possible $Q^2$ values and the maximum CLAS torus magnetic field (current = $3375\;\rm{A}$), which allowed
us to achieve the best acceptance and resolution for out-bending charged particles including the backward-angle $\pi^+$s.
In this analysis, we detected the scattered electron and the final state pion in CLAS.
The $\theta$ coverage in polar angle ranges from about $8^{\circ}$ to $140^{\circ}$ for $\pi^+$.
The exclusivity of the $e p \to e^\prime n \pi^+$ reaction was established by making a cut around the neutron mass
in the missing mass $M_X$ spectrum of the $e p \to e^\prime \pi^+ X$ system.
Details of the data analysis are given in Ref.~{\cite{KPARK13}} where the same data
set and $e p \to e^\prime n \pi^+$ process was analyzed to extract GPDs, in that case focusing
on the forward-angle pions. 

Although the kinematics of the particles was a bit different in the present analysis,
the general particle identification procedures and the data analysis techniques are the same as in Ref.~\cite{KPARK13}.
Therefore, in the following, we sketch just the main steps of the present data analysis.
The CLAS electron trigger required a minimum energy in the EC
in coincidence with a CC signal. To improve the electron selection, additional cuts were applied on the EC energy, corresponding to a minimum electron momentum of 0.64 GeV. A $z$-vertex cut (-80 mm $< z_{vtx} <$ -8 mm, target center was at -40 mm)
was made around the target location.
A cut on the number of photo-electrons in the CC and general geometric fiducial volume cuts were made in
order to keep only regions of uniform detector efficiency, which could reliably be reproduced by our Monte-Carlo software/program.
Pions were identified by a coincidence of
signals in the DC and TOF counters and by the time-of-flight technique within the fiducial cut regions.
Missing TOF channels and bad DC regions were excluded from the analysis. All cuts were applied to both experimental
and simulated data. Ad-hoc kinematic corrections were used to improve the measured angles and momenta of the particles
due to misalignments of CLAS sectors or magnetic field inhomogeneities~{\cite{KPark00}.
 \begin{table}[htb]
 \begin{center}
 \caption{Kinematic Bins}
 \begin{tabular}{cccc}
 \hline
 Variable & Number of bins & Range & Bin size \\
 \hline
 \hline
 $W$  &   1  & $2.0 - 2.4$ GeV  & $400$ MeV\\

 $Q^2$ &  6 & $1.6 - 4.5$ $\rm{GeV}^2$ & varying \\

 $\Delta_T^2$  & 1 & $0 - 0.5$ $\rm{GeV}^2$& 0.5 $\rm{GeV}^2$\\

 $\varphi^*_{\pi}$ & 9 & 0$^\circ$ - 360$^\circ$ & 40$^\circ$ \\
 \hline
 \end{tabular}
 \label{tab:kine_range}
 \end{center}
 \end{table}
 \begin{figure}[htb]
   \begin{center}
\includegraphics[angle=0,width=6.9cm,height=6.4cm]{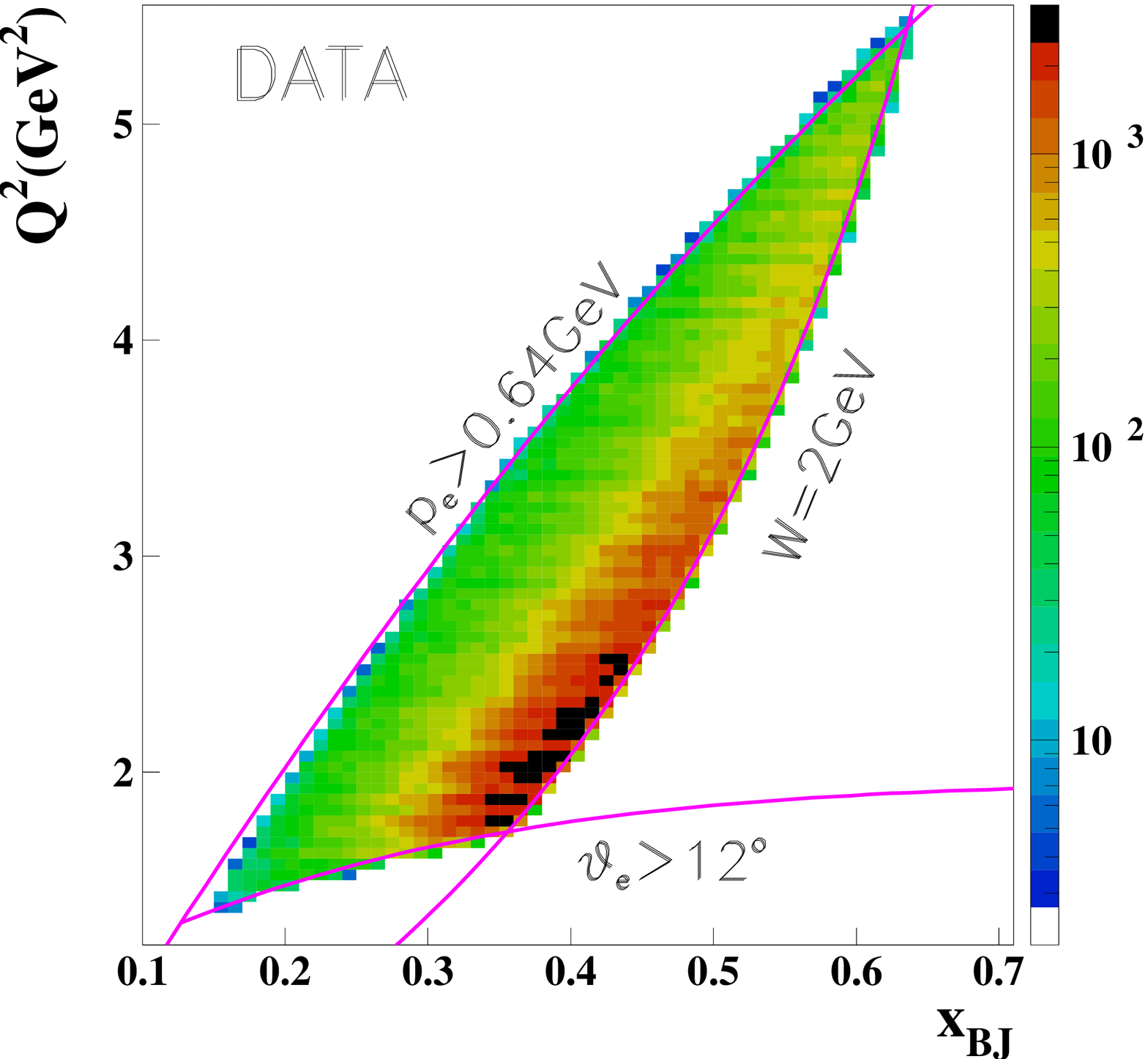}\\
\includegraphics[angle=0,width=6.9cm,height=6.4cm]{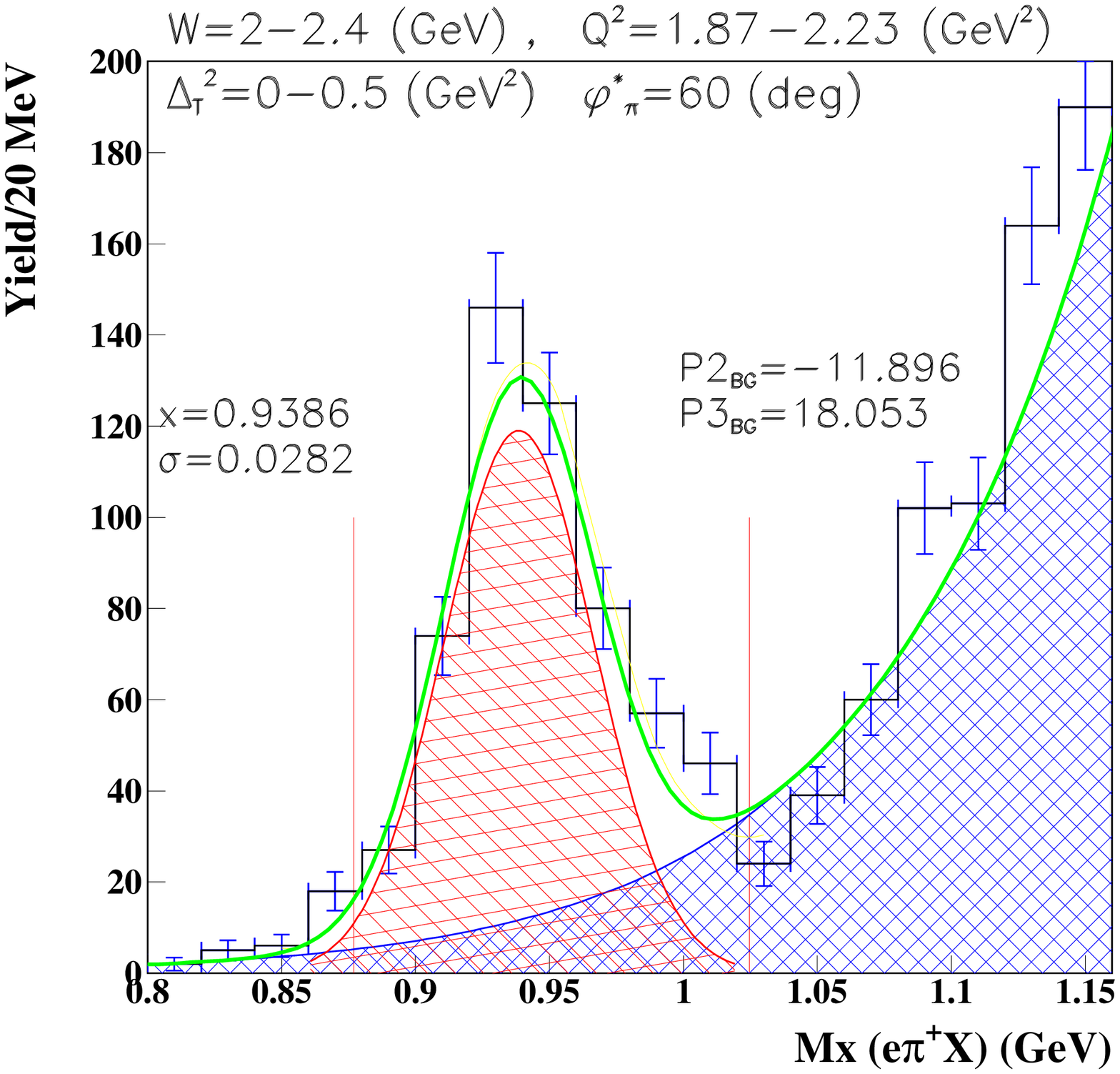}
        \caption[Kinematic coverage in this analysis]{
          (Color online) \textbf{Top}: kinematic coverage in $Q^2$ versus $x_{BJ}$. \textbf{Bottom}: an example of the neutron missing mass peak fit. Here $\langle W\rangle$=2.2 GeV, $\langle Q^2 \rangle$=2.05 GeV$^2$, $\langle \Delta_T^2 \rangle = 0.25$ and $\varphi_{\pi}^*=60$ (deg). The red shaded curve (a skewed Gaussian fit) is the signal + radiative tail. The blue shaded curve (exponential+polynomial fit) is the background and the green curve is the sum of both signal and background. Neutrons were selected from the region between the vertical lines.
        }
        \label{fig:kinebin}
   \end{center}
\end{figure}

The top plot of Fig.~\ref{fig:kinebin} shows the kinematic coverage of the data in $Q^2$ and $x_{BJ}$
after all electron cuts. Two additional cuts, $\Delta_T^2 < 0.5\;\rm{GeV^2}$ and $\cos\theta^*_{\pi} < 0$, selected
backward-angle pions, applicable to the TDA formalism.
We binned our phase space trying to keep roughly equal statistics in each bin. Table~\ref{tab:kine_range} shows the kinematic bins
used in this analysis.
The bottom plot of Fig.~\ref{fig:kinebin} shows a typical missing mass $M_X$ spectrum. The background under the neutron missing-mass peak was due to particle misidentification and/or multi-pion channels, smeared by the experimental resolution.
This background was estimated by a Gaussian fit to the neutron peak plus an exponential background.
Several functions were tested to fit the data. The variation among these fits resulted in a 4\% systematic uncertainty. After subtraction, the resulting neutron peak (position and resolution) in the data agreed
with the Monte Carlo simulation.
The Monte-Carlo software, GSIM, was based on GEANT3 and it is the standard software simulation package for CLAS data analysis.
Simulated data go through the same chain of reconstruction codes as real data. Tunable parameters for each detector
were adjusted so that the Monte-Carlo distributions matched the experimental data.
We used a phase-space-based event generator to simulate $e p \to e^\prime n \pi^+$~{\cite{fsgen}}
with the addition of an exponential $e^{Au}$-dependence with an ad-hoc parameter $A$ to reproduce the pion angular dependence at large angles.
The determination of CLAS acceptance and efficiency was done for
each four-dimensional bin. The ratio between the number of generated and reconstructed events in a bin,
after taking into account all cuts and corrections, was applied as a correction factor.
Approximately $300$ million $e p \to e^\prime n \pi^+$ events were generated in the kinematic range of Table~\ref{tab:kine_range}.
Radiative corrections were applied
using the extended ExcluRad~{\cite{afanasev}} program.

We have extracted the $\sigma_T + \epsilon \sigma_L$ (=$\sigma_U$),
$\sigma_{LT}$ and $\sigma_{TT}$ cross sections as a function of $Q^2$ at a given $W$ and $-u$ kinematics.
The structure functions $\sigma_U$, $\sigma_{LT}$ and $\sigma_{TT}$ from the experimental data
were fed into the program, and the ratio of the computed cross sections, with radiation on and off, were generated for each bin. The systematic uncertainties associated with this correction were determined using different parameters of the program. This resulted in a 10\% systematic uncertainty,
which turned out to be the dominant contribution compared to the other systematic uncertainties.
The cut values, bin sizes, and fitting functions were varied in order
to test the stability of our final cross sections. The systematic uncertainty associated 
with electron identification was estimated to be less than $2\%$. For the $\pi^+$ identification, the systematic uncertainty is negligible.
A one-$\sigma$ change in the neutron missing mass cut yields an average $3\%$ systematic uncertainty.
The $\Delta_T^2$ cut was changed between 0.5 $\rm{GeV^2}$ and 1.0 $\rm{GeV^2}$, resulting in $< 1 \%$ uncertainty.
Due to the limited statistics of the experimental data, we used $9$ bins in $\varphi_{\pi}^*$. We tested an analysis
with $12$ bins in $\varphi_{\pi}^*$, which resulted in a variation of $4\%$. The uncertainties associated with the luminosity
and the density and length of the target were estimated to be 2\% and 1\%, respectively.
The total systematic uncertainty was estimated to be $12\%$.
%
%

We extracted the $\varphi_{\pi}^*$-dependent cross sections of the $e p \to e^\prime n \pi^+$ reaction at the average kinematics
$\langle W \rangle$= 2.2 GeV and $\langle -u \rangle$= 0.5 GeV$^2$, for six different $Q^2$ values: 1.71, 2.05, 2.44, 2.92, 3.48 and 4.16 GeV$^2$. The data points are included in the CLAS Physics Database~\cite{CLASDB}.
This covers $\xi$ in the range $[0.1 - 0.45]$. Fig.~\ref{fig:phi-crs00} shows these results.
The differential cross sections are fit to Eq.~(\ref{eq:reaction02}) taking only statistical uncertainties into account.
The average $\chi^2$ per degree of freedom of the five lowest-$Q^2$-bin fits was $\sim 2.6$ except $Q^2$= 4.16 GeV$^2$ due to lack of data. 
Since the CLAS acceptance showed a complicated $\varphi_\pi^*$-dependence around $\varphi_\pi^* \sim 0$,
we took into account an additional systematic uncertainty of $\varphi_\pi^*$ binning in the acceptance calculation for extraction
of the structure function.
\begin{figure}[!htb]
\begin{center}
        \includegraphics[angle=0,width=0.47\textwidth,height=100mm]{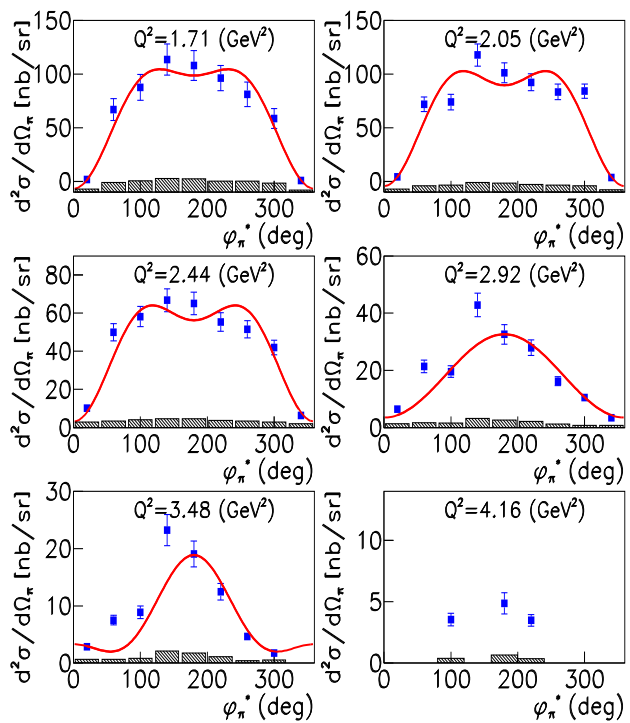}
\caption[$\varphi_{\pi}^*$ dependent ${d\sigma}/{d\Omega_{\pi}^*}$ cross sections.]{ (Color online) 
          $\varphi_{\pi}^*$-dependent differential cross sections (${d\sigma}/{d\Omega_{\pi}^*}$) for  $Q^2=1.71$, $2.05$, $2.44$, $2.92$, $3.48$, and $4.16\;\rm{GeV^2}$ in the backward region. The red solid curves show the full fit results using Eq.~(\ref{eq:reaction02}). The shaded areas show the systematic uncertainty. The Y-axis in the lowest two $Q^2$ bins has negative offset to show full fit range.
          \label{fig:phi-crs00}
        }
\end{center}
\end{figure}

Figure~\ref{fig:stf10} shows the $Q^2$-dependence of $\sigma_U$, $\sigma_{LT}$ and $\sigma_{TT}$, obtained at the average kinematics $\langle W \rangle$= 2.2 GeV and $\langle -u \rangle$= 0.5 GeV$^2$. We note that all three cross sections have a strong $Q^2$-dependence.
The TDA formalism predicts the dominance at large $Q^2$ of the transverse amplitude.
Therefore, in order to be able to claim the validity of the TDA approach, it
is necessary to separate $\sigma_T$ from $\sigma_L$ and check
that $\sigma_T \gg \sigma_L, \,\sigma_{TT}$ and $\sigma_{LT}$.
With only this set of data at fixed beam energy, we cannot do the experimental separation of $\sigma_T$ and $\sigma_L$.
However, we observe that $\sigma_{TT}$ and $\sigma_{LT}$ are roughly equal in magnitude and have a similar $Q^2$-dependence.
Their significant size (about 50\% of $\sigma_U$) implies an important contribution of the transverse amplitude in the cross section.
The theoretical TDA description of $\sigma_{TT}$ and $\sigma_{LT}$ yields a suppression factor of order $\Delta_T^2 / Q^2$ with respect to $\sigma_T$. 
In Fig.~\ref{fig:stf10}, we compare our data for
$\sigma_U$ to the theoretical predictions of $\sigma_T$
from the nucleon pole exchange $\pi N$ TDA model suggested in Ref.~\cite{Pire:2011xv}.
The curves show the results of three  theoretical calculations
using different input phenomenological solutions for the nucleon DAs
with their uncertainties represented by the bands.
Black band: BLW NNLO~\cite{BLWNNLO}, dark blue band: COZ~\cite{COZ}, and light blue band: KS~\cite{KS}.

\begin{figure}[!htb]
\begin{center}
  \includegraphics[angle=0, width=8.4cm]{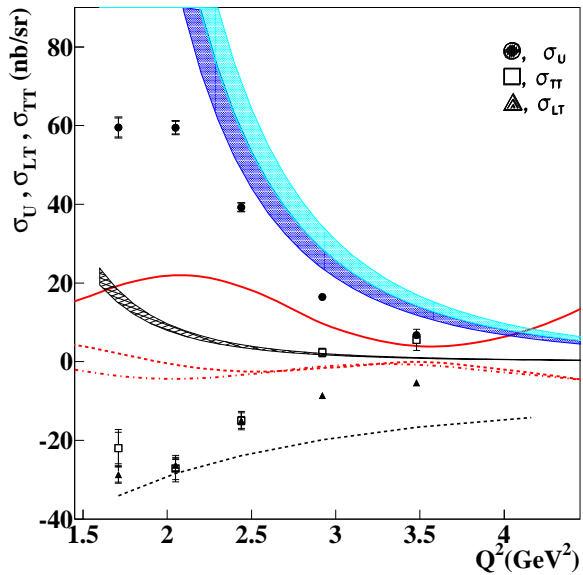}
        \caption[The structure functions in terms of $Q^2$.]{
 	  (Color online) The structure functions $\sigma_U$ ($\bullet$), $\sigma_{TT}$ ({$\square$}) and  $\sigma_{LT}$ ({$\blacktriangle$}) as a function of $Q^2$. The inner error bars are statistical and the outer error bars are total (=$\sqrt{\delta^2_{stat}+\delta^2_{sys}}$) uncertainties. The curves are explained in text.
        }\label{fig:stf10}
\end{center}
\end{figure}

The other curves (bold red solid:~$\sigma_U$, dashed:~$\sigma_{LT}$, dot-dashed:~$\sigma_{TT}$) are the predictions of the effective hadronic description of Ref.~\cite{Laget:2009hs}, which is based on the exchange of $\pi$- and $\rho$-Regge trajectories in the $t$-channel, $N$- and $\Delta$-Regge trajectories in the $u$-channel and unitarized $\pi$ and $\rho$ rescattering. It reproduces the high energy ($\sqrt{s}=4$ GeV) SLAC~\cite{RLAnderson} photoproduction data fairly well. When supplemented with $t$-dependent electromagnetic form factors, according to the prescription of Ref.~\cite{JMLaget2004}, it also reproduces the HERMES~\cite{HERMES} electroproduction data ($\sqrt{s}=4$ GeV and $Q^2$=2.4 GeV$^2$). At lower energies ($\sqrt{s}$=2.2 to 2.5 GeV), this leads to a fair accounting of the published JLab data~\cite{KPARK13} at low and intermediate $t$. The model is close to the data at high $Q^2$ but misses them at lower $Q^2$. The black dashed curve shows  $ (- \Delta_T^2 / Q^2)\sigma_U$ parameterized from the experimental data.
The approach to scaling in backward electroproduction was also studied in the $ep \to ep\gamma$ and $ep \to ep\pi^0$ reactions at $Q^2=1$ GeV$^2$ and $W^2 < 4$ GeV$^2$ in Ref.~\cite{Laveissiere2004gr}.

In summary, we have measured for the first time the cross section of $e p \to e^\prime n \pi^+$ at large photon virtuality,
above the resonance region, for pions at backward angles, using the CLAS detector at Jefferson Lab. 
The motivation to address 
such a kinematic regime was
provided by the potentially applicable collinear factorized description in terms of
nucleon-to-pion TDAs that encode valuable nucleon structural information. 
The final goal was an experimental validation 
of the factorized description and the extraction of nucleon-to-pion TDAs
from the observed quantities. 
Our analysis represents a first encouraging step towards this goal. 
We see a very reasonable agreement between the TDA model-dependent calculation and our data.
However, this is not incontrovertible evidence for the 
validity of the factorized description, 
since the TDA-based description 
and the phenomenological Regge-pole exchange model of 
Laget~\cite{Laget:2009hs} yield similar results.
From theory, there exists several signs of the onset of 
factorization. The most obvious ones are  the characteristic scaling 
behavior of the cross section in $1/Q^8$ 
and the related twist counting rules that lead to the dominance of the transverse 
polarization of the virtual photon, which results in 
$\sigma_T \gg \sigma_L, \, \sigma_{LT}$ and $\sigma_{TT}$.
Such experimental tests require both the explicit 
separation of $\sigma_T$ and $\sigma_L$ 
and the precise cross section measurements over a
wide range of $Q^2$ 
to provide a large lever arm for the $1/Q$-scaling tests. 
Another way to confirm the validity of the factorized description
is to use a polarized target to measure the appropriate spin observables. 
For example the transverse single spin asymmetry (TSSA)~\cite{Lansberg:2010mf} 
is sensitive to the imaginary part of the reaction amplitude.
The considerable size of the TSSA can be most easily interpreted as a sign of the
validity of the TDA-based approach.    
Additional evidence for the TDA-based description can be provided by observing the universality of
the nucleon-to-pion TDA accessed in other reactions, 
which can be studied at \={P}ANDA@GSI-FAIR~\cite{Lansberg:2012ha,Pire:2013jva,Singh:2014pfv,Singh:2016qjg} J-PARC~\cite{Pire:2016gut} as well as a
variety of light meson electroproduction reactions ($\eta$, $\eta^\prime$, $\rho$, $\omega$) at JLab~\cite{Pire:2015_prd96}. 

%
%

We acknowledge the outstanding efforts of the staff of the
Accelerator and the Physics Divisions at Jefferson Lab in making this experiment possible. This work was supported in
part by the US Department of Energy, the National Science
Foundation (NSF), the Italian Istituto Nazionale di Fisica Nucleare (INFN),
the French Centre National de la Recherche Scientifique (CNRS), the
French Commissariat {\`a} l'Energie Atomique, the UK's Science and Technology Facilities Council, and the National Research Foundation (NRF) of Korea.
The Southeastern Universities
Research Association (SURA) operated the Thomas
Jefferson National Accelerator Facility for the US Department
of Energy under Contract No.DE-AC05-06OR23177.
%
%
%


\end{document}